\documentclass{aa}
\usepackage{graphicx}
\usepackage{natbib}
\bibpunct{(}{)}{;}{a}{}{;}
\begin{document}

\headnote{Research Note} 
\title{The Effect of Scattering on Pulsar Polarization Angle} 
\authorrunning{X.H. Li \& J.L. Han}
\titlerunning{the Effect of Scattering on Pulsar Polarization Angle} 
\author{X.H. Li \and J.L. Han}

\offprints{hjl@bao.ac.cn (J.L. Han)}

\institute{National Astronomical Observatories, Chinese Academy of Sciences, 
        Jia-20 DaTun Road, Chaoyang District, Beijing 100012, China}

\date{Received 2003 May 02 / accepted 2003 July 29}

\abstract{
The low-frequency profiles of some pulsars manifest temporal 
broadening due to scattering, usually accompanied by flat 
polarization position angle (PA) curves. Assuming that the 
scattering works on the 4 Stokes parameters in the same way,
we have simulated the effect of scattering on polarization profiles
and find that the scattering can indeed flatten the PA curves. 
Since the higher-frequency profiles
suffer less from scattering, 
they are convolved with scattering models 
to fit the observed low-frequency profiles. 
The calculated flat PA curves exactly reproduce 
the corresponding observations. 
}

\maketitle

\section{Introduction}
The polarization properties of more and more pulsars have been observed. 
They are important for our understanding of the pulsar emission process.
The $S$-type PA curves of most pulsars are related to the rotation
of the magnetic field plane where emission is generated, as described 
by the rotating-vector model. Recently, we noticed that if good 
quality measurements are available, almost all scatter broadened 
profiles have a flat PA curve in the trailing part without exception. 
Most previous work on scattering has concentrated on the total 
intensity profile \citep[e.g.][]{r77,lg98,ki93,bcc03}. Only a few 
just mention the possible effect of scattering on PA curves 
\citep{gil85,wxq02} except for the early work by \citet{k72} on 
the Vela pulsar. In this research 
note, we discuss the scattering models and the effect of scattering
on both total and polarized intensity profiles.

Scatter broadening is related to the geometry of the scattering disk
\citep[see ][]{lg98}. Compared to the radiation coming from the center
of the disk, the extra delay $\tau$ of the scattered rays from a radius 
$r$ in a disk at distance $L$ should be $\tau=r^2/2cL$. Considering
the probability of a scattered ray from $r$ to $r+dr$, one finds an
intensity variation with delay $t$: $I(t)\propto \exp(-2cL/r^2 \;t)$.
Whenever the polarized radiation passes through the 
interstellar medium, as shown by \citet{mm00}, all Stokes parameters
should scintillate like the total intensity. Therefore, it is natural
to assume that the scattering process works similarly on all Stokes 
parameters. We simulated the scattering effect on pulsar Stokes profiles
and then checked the observational data of several pulsars. Using the 
observed high frequency mean pulse profiles without obvious scatter
broadening as the intrinsic profiles, we can recover 
the low frequency observed profiles, including the flat PA curves, 
very well by convolving them with scattering 
models. We note that \citet{k72} had also shown that the broadened 
polarization profiles of the Vela pulsar (PSR B0833$-$45) are consistent
with the scattering model.

\begin{figure*}[ht]
\centering
    \includegraphics[scale=0.6,angle=-90]{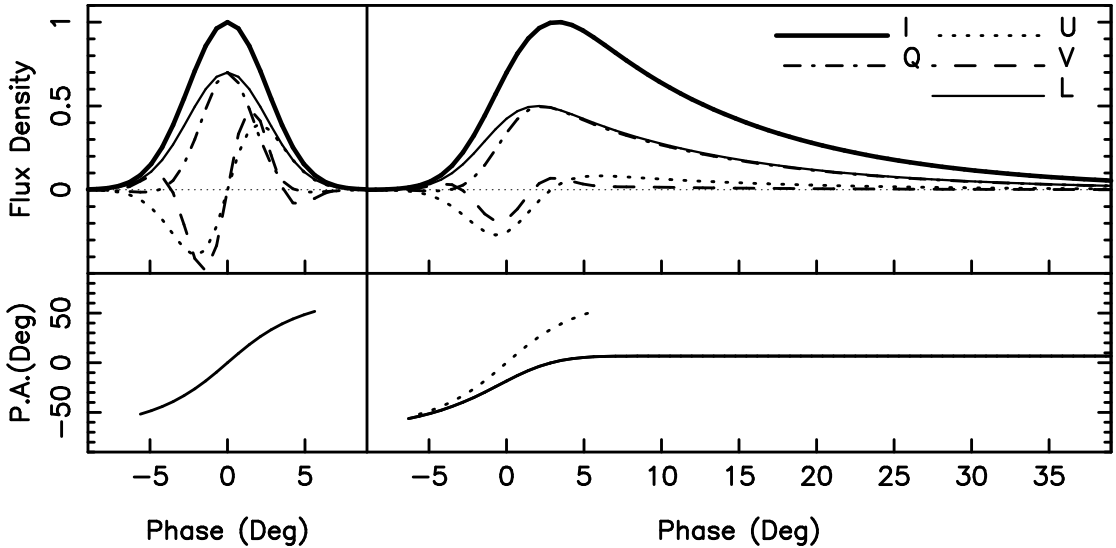}
        \caption{ 
The scattering effect of a thin screen on simulated Stokes polarization 
profiles. The simulated intrinsic $I$, $Q$, $U$, $V$, $L$ and $PA$ are 
shown in the left panels, and the scattered ones in the right panels. 
The dotted line in the right-lower panel is the original PA. Obviously 
the PA curve of scattered profile is very flatened in the long pulse
tail.
}
\end{figure*}

\section{Scattering Model and Simulation}
The scattered pulse $y(t)$ can be derived by convolving the intrinsic 
pulse $x(t)$ with the response scattering function $g(t)$:
\begin{equation}
y(t)=\int x(\zeta)g(t-\zeta)d\zeta.
\label{convolvtion}
\end{equation}
The thin screen of interstellar medium between the pulsar and the 
observer can be modeled by a response function $g(t)=\exp(-t/\tau_{\rm sc})$ 
for $t>0$ \citep{ki93,w72}. The scattering time scale $\rm \tau_{\rm sc}$ can 
be related to the dispersion measure (DM) and the observation wavelength
($\lambda$) by an empirical relation
\begin{equation}
\tau_{\rm sc}=4.5\times10^{-5}DM^{1.6}\times(1+3.1\times10^{-5}DM^3)\lambda^{4.4}.
\label{timescale}
\end{equation}
with $\rm \tau_{\rm sc}$ in millisecond, DM in $\rm pc\,cm^{-3}$ and 
$\lambda$ in meter \citep[see][]{mr01}.

The total intensity ($I$) profile of the simulated pulse has a Gaussian 
shape, with linear polarization ($L$) degree of 70\% and a PA ($\psi$)
following the rotating-vector model; the Stokes parameters $Q$ and $U$
can be written as $Q=0.7I\cos(2\psi)$, $U=0.7I\sin(2\psi)$. 
The circular polarization ($V$) is assumed to have a sense reversal 
(see the left panels in Fig.~1). If the scattering acts on all the 
Stokes parameters as described by Eq.~(1), we can convolve $I$, $Q$,
$U$, $V$ with the thin scattering model to get all scattered quantities.
Simulation results are shown in Fig.~1. We can see in the right 
panels that the PA curve of the scattered profile is flattened and
the circular polarization almost vanishes in the trailing part.

\section{Scattering Effect on Observed Pulsar Profiles}

To check the effect of scattering on polarization profiles,
we downloaded high quality multifrequency polarization data of 
5 pulsars \citep{gl98} available from EPN database\footnote 
{http://www.mpifr-bonn.mpg.de/pulsar/data/}\citep{ljs+98}, for which 
scatter broadening is very obvious in low frequency 
profiles (see Fig.~2-6).

The high frequency profiles without obvious scatter broadening 
were considered to be the intrinsic emitted pulse profiles 
for reference. The scattering time scales (Col.~7, 8 \& 9
in Table~1)
at lower frequencies can be either estimated using Eq.~(2) in Col.~7 
or derived by maximizing the correlation coefficients between the 
total intensity
profile observed at lower frequency and the profiles convolved from
high frequency profiles (Table~1) with a thin screen model (Col.~9).
In fact,
it can also be estimated directly from the exponential tails (Col.~8).
Using the
scattering time scale from the cross-correlation (with an uncertainty of 
about 20\%), we then calculated each Stokes profile of 4 pulsars 
using the thin screen model.  The convolved and  observed profiles 
as shown in Fig.~2 and Fig.~4-6 are very consistent. We noticed that 
the convolved 
profiles of PSR  B1838$-$04 at 0.606 GHz using a thin screen model do
not fit the observed ones (Fig.~3), so we tried a thick screen and an 
extended medium \citep{w72} for response functions.  The extended medium 
model
\begin{equation}
g(t)=(\frac{\pi^{5}\tau_{\rm sc}^3}{8t^5})^{1/2}\exp(-\frac{\pi^2 \tau_{\rm sc}}{4t})
\label{extendedmedium}
\end{equation}
does work very well.

\begin{table*}[!htbp]
\caption{ Parameters of pulsar polarization profiles with obvious scattering
at the lower frequency. Columns are pulsar name, period, 
dispersion measure (DM), the higher and lower observation frequencies,
sampling time at the lower frequency, the scattering time scale calculated
by the empirical relation given in Eq.~(2) ($\rm \tau_{em}$),
the time-scale of exponentially decaying tails ($\rm \tau_{ex}$),
the scattering time scale estimated from real data deconvolution ($\rm \tau_{\rm sc}$)
and the corresponding scattering model used to fit the low frequency profiles.}
\label{tab:}
\begin{tabular}{lclcccccrc}
\noalign{\smallskip}
\hline
\hline
\noalign{\smallskip}
PSR & P&DM&Freq.&Freq.&$\rm T_{s}$&$\rm \tau_{em}$&$\rm \tau_{ex}$&$\rm \tau_{\rm sc}$&Scattering Model\\
B-name & s&$\rm pc\,cm^{-3}$&GHz &GHz &ms &ms &ms&ms& \\
\hline
\noalign{\smallskip}
1831$-$03& 0.6867&  235.8&0.610&0.408&  3.60& 29.54& 19.43 &18.65&thin screen\\
1838$-$04& 0.1861&  324.0&1.408&0.606&  3.02& 22.32& 20.45 &12.33/24.54&extended/thin\\
1841$-$05& 0.2557&  411.0&1.408&0.610&  3.84& 57.13& 53.74 &61.28&thin screen\\
1859$+$03& 0.6554&  402.9&0.925&0.606&  3.75& 60.79& 14.77 &13.74&thin screen\\
1946$+$35& 0.7173&  129.1&0.610&0.408&  1.97&  1.87& 16.00 &12.74&thin screen\\
\noalign{\smallskip}
\hline
\hline
\noalign{\smallskip}
\end{tabular}
\end{table*}

\begin{figure}
    \includegraphics[scale=0.48,angle=-90]{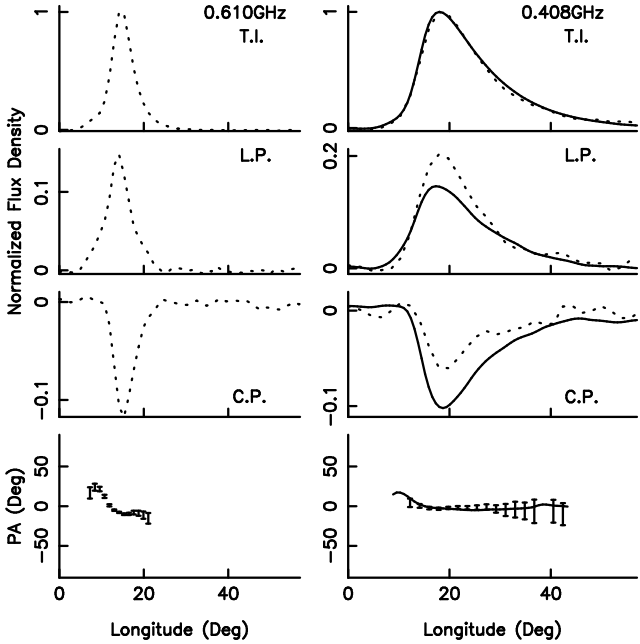}
        \caption{
Polarization profiles of PSR B1831$-$03. Total intensity (T.I.), 
linearly polarized intensity (L.P.), circular polarization (C.P.)
and polarization angle (P.A.) are shown in different panels for
two frequencies. All intensity profiles have been normalized 
to the peak of total intensity.
The dotted lines are observed data. The errors of PA were given by
$\sigma_{\psi}=\sqrt{Q^2\sigma_U^2+U^2\sigma_Q^2}/(2L^2)$. 
The solid lines are profiles convolved by the highest frequency
pulse profile with a thin scattering screen. 
}
        \label{B1831-03}
\end{figure}

\begin{figure}
    \includegraphics[scale=0.48,angle=-90]{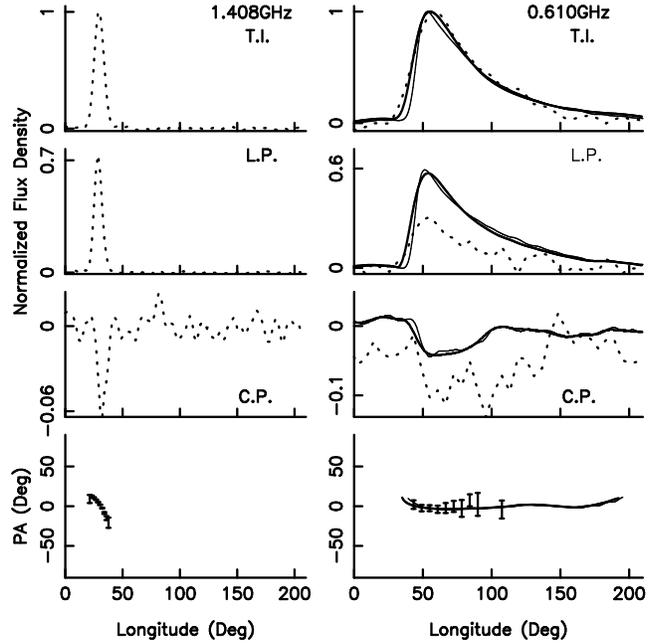} 
        \caption{The same as Fig.~2, but for PSR B1838$-$04 and convolved 
with an extended scattering medium (thick lines) and a thin model (thin lines).}
        \label{B1838-04}
\end{figure}

\begin{figure}
    \includegraphics[scale=0.48, angle=-90]{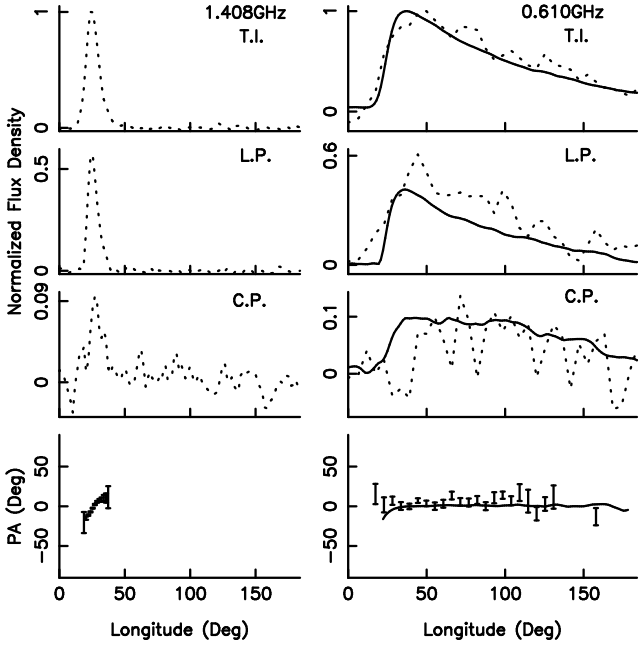} 
        \caption{The same as Fig.~2, but for PSR B1841$-$05.}
        \label{B1841-05}
\end{figure}

\begin{figure}
  \includegraphics[scale=0.48,angle=-90]{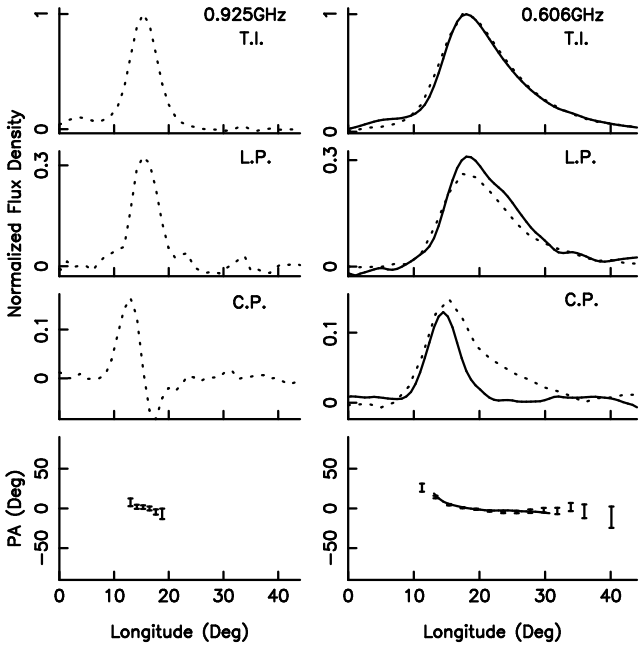} 
        \caption{The same as Fig.~2, but for PSR B1859+03.}
        \label{B1859+03}
\end{figure}

\begin{figure}[!ht]
    \includegraphics[scale=0.48,angle=-90]{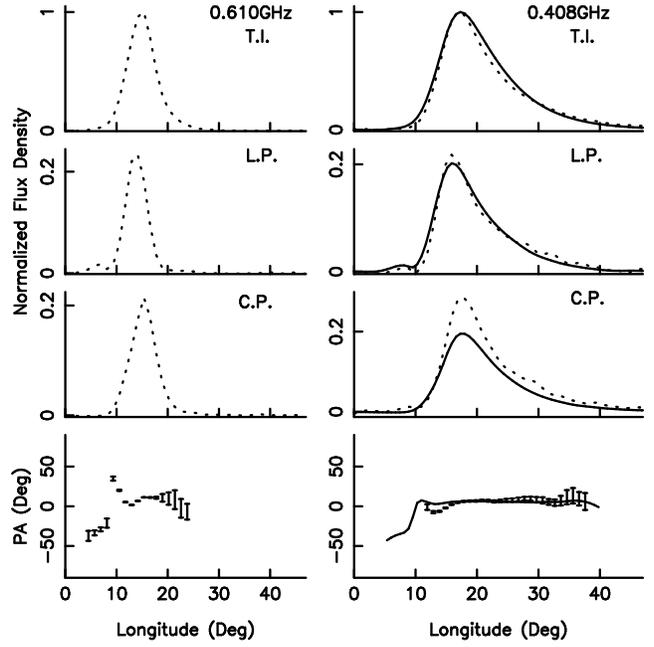} 
        \caption{The same as Fig.~2, but for PSR B1946+35.}
        \label{B1846+35}
\end{figure}

 The scattering time-scales we deconvolved from lower frequency
data are consistent with earlier estimates if scaled with frequency,
i.e. PSR B1831$-$03 in \citet{rmd+97} and others in \citet{tml93}.
While the estimate given by the empirical relation in Eq.(2) 
is a good description of the general tendency against DM, for a 
given DM they can differ by more than one magnitude \citep{mr01,rmd+97}. 

Obviously when $\rm \tau_{\rm sc}$ increases, the PA curves can be
completely flattened irrespective of the original PA curve shape.
In this case, the maximum gradient of the PA curve has been
diminished and cannot be directly used to determine the 
geometry of the pulsar emission region \citep[e.g.][]{wxq02}. As we
have tested, it is in principle possible to recover all Stokes
profiles from the scattered ones as long as the a proper scattering
model is used. However, one cannot be sure that the scattering 
material is of the thin-screen type, or extended, or in 
multi-thin-screen form. Consequently, the merit of deconvolved profiles 
is difficult to assess. For example, two possible results for PSR 
J1852+0031 from the two kinds of scattering medium given by \citet{bcc03}
are both acceptable. It would be difficult to say which one is more likely
to be the intrinsic one.

The recent Parkes multibeam Survey has discovered many pulsars with 
scatter broadened profiles \citep{Man01}. Polarization observations
of some of them have shown the flattened PA curves in the trailing part
\citep[e.g. PSR J1730$-$3350 in][]{cmk2001}. The leading part in fact
cannot be used to determine the geometrical parameters of the pulsar, 
mainly because it has also been influenced by scattering (see simulation 
in Fig.~1).  However, if the scattering model can be determined by 
comparing the scattered profile with one at higher frequency, then
it is worth trying to recover the intrinsic Stokes profiles.

It is also worthy to note that not all flat PA curves are related 
to the scattering effect. For example,  the flat PA curves of
millisecond pulsars, e.g. PSR B1937+21 \citep{ts90}, at higher 
frequencies are obviously intrinsic, and probably not related 
to the scattering process.

From simulations, we also noticed that when the sense reversal 
appears in the intrinsic circular polarization profile, the scattering
can largely weaken the circular polarization in the trailing part,
as shown by PSR B1859+03 (Fig.~\ref{B1859+03}). If there is no sense
reversal, as in the other 4 pulsars shown, the single-hand circular 
polarization profiles are simply stretched.

\section{Conclusion}
We investigated the scattering effect on polarization profiles, 
especially position angle.
The scattering can flatten the PA curves.
Taking the high frequency profiles of 5 pulsars as intrinsic pulses, 
the observed low frequency scattered polarization 
profiles can be reproduced by convolving the high frequency observations
with the scattering models. 

\section*{Acknowledgments}
We thank the referee, Prof. R.G. Strom, for his insightful suggestions and
careful readings which improved the paper a lot. We are very grateful to
X.H. Sun for carefully reading the manuscript. JLH is
supported by the National Natural Science Foundation of China (19903003 and
10025313) and the National Key Basic Research Science Foundation of China
(G19990754) as well as the partner group of MPIfR at NAOC.

\end{document}